\documentstyle[11pt,newpasp,twoside,epsf]{article}
\def\edcomment#1{\iffalse\marginpar{\raggedright\sl#1\/}\else\relax\fi}
\reversemarginpar

\begin{document}
\title{Towards Population III: Simulations of Primordial Gas Collapse
and Fragmentation}
\author{Paolo Coppi}
\affil{Department of Astronomy, Yale University, P.O. Box 201801,
New Haven, CT 06520-8101, USA}
\author{Volker Bromm}
\affil{Institute of Astronomy, Cambridge University, Madingley Road, 
Cambridge CB3 0HA, UK }
\author{Richard Larson}
\affil{Department of Astronomy, Yale University, P.O. Box 201801,
New Haven, CT 06520-8101, USA}

\begin{abstract}
We briefly review the motivations for studying the formation of
the first ``Population III'' stars and present recent
results from our numerical simulations in this area. We discuss
the new questions  raised as a result
of the simulations presented by us and others at this 
meeting.
\end{abstract}

\section{Introduction: Why Consider Population III (Again)?}

At first glance, 
the study of the very first stars in the universe
might appear to be a rather academic and even quixotic endeavor.
After all, we have never directly seen such stars.  
Since we believe that heavy elements
are synthesized almost exclusively in stars, the very
first ``Population III'' stars were, by definition,  made of zero metal gas.
Intensive searches for metal poor stars in the halo galaxy, however, 
have only turned up stars with  metallicities $Z \ga 10^{-4} Z_{\odot}$
(Beers 2000). Furthermore, the formation of a zero metal star is in some
sense a rather singular event. Once that star goes supernova
and pollutes its environment with metals, the stars that 
subsequently form in its vicinity will no longer
be Population III (with zero metallicity).  Population III
may also be suicidal in the sense that UV radiation from
the first stars could destroy
the molecular hydrogen that allows primordial gas to cool
and form these stars in the first place.
There has therefore been considerable speculation, e.g., Cayrel (1986)
and Haiman, Rees, \& Loeb (1997),
that the Population III (Pop. III) phase in the evolution
of our universe was very brief indeed.

Nonetheless, while the Pop. III phase may have been a brief
one, it seems to have been a pervasive one.  Even in the
least evolved regions of the universe that we can
probe today, the high redshift  Lyman $\alpha$ clouds,
we still
find evidence for a non-zero metallicity $Z_{min}\sim 10^{-3}Z_{\odot}$ 
(Cowie \& Songaila 1998). More importantly, the epoch of the Pop. III 
stars probably  represents the first  substantial input of energy,
photons, and metals into the universe since the Big Bang
and marks the end of the ``dark ages'' 
(e.g., Loeb 1999, Rees 1999) that started when
the cosmic background light redshifted out of the visible range
(at $z\sim 1000$). Because structure formation depends critically 
on the ability of baryons to cool, and this in turn depends
critically on the metallicity 
and ionization state of the baryonic gas, the ``feedback'' effects
of the first stars on the intergalactic medium (IGM)
in fact play a subtle but key role in determining
the subsequent evolution of the Universe. The significant interest
in primordial star formation at this conference is therefore
not an accident.  Large-scale cosmological
simulations (e.g., Ostriker \& Gnedin 1996; 
Abel et al. 1998;  Fuller \& Couchman 2000)
are just now  becoming good enough  to resolve
the $\sim 10^6 M_\odot$ scales of  the first objects
in the universe that can cool and collapse (Tegmark et al. 1997).
Understanding the star formation that will occur in these objects
and accurately incorporating its  effects into cosmological codes
therefore represents the next major challenge 
in our quest to integrate forwards from the Big Bang.

The interest in Population III, however,  is not purely a theoretical
one.  Studies of cosmic microwave background (CMB) anisotropies 
are now providing us with information on the physical conditions 
at $z\sim 1000,$ when the universe became transparent to CMB 
radiation, while  
observations of high-redshift quasars and galaxies tell us about the 
universe at redshifts $z \sim 5-6$ below.
Current observations, however, give us little information
on  the ``dark ages,'' the crucial epoch in between.
Reaching into this epoch is thus
our next observational challenge
and is the primary motivation, for example,
of the {\it Next Generation
Space Telescope} (NGST) which will provide unprecedented 
sensitivity at near-infrared wavelengths (Loeb 1998). 
The study of the first stars is thus timely,
providing a theoretical framework for the interpretation of what NGST
might discover, less than a decade from now.          
Even if NGST does not directly image the first stars,
it will probe the epoch of the reionization of the IGM (e.g., Barkana,
this proceedings). UV photons from the first stars, perhaps
together with an early population of quasars, may have contributed 
significantly to this reionization (e.g., see
 Ciardi et al. 2000; Miralda-Escud\'{e}, 
Haehnelt, \& Rees 2000). If the reionization  occurred early
enough, CMB fluctuations on scales $\la 10^\circ$ will 
be damped by electron scattering (e.g,. Haiman \& Loeb 1997)
and this effect could be detected by other next generation
instruments like MAP and PLANCK. The energy input from the first
stars may also have left a small but measurable imprint on the CMB 
on very small scales (e.g., see the contributions 
here by Sugiyama et al. and Bruscoli et al.).
In sum, the implications of Pop.III star formation might be 
testable in the not too distant future!

Before that day arrives, however, a skeptical observer might still
question whether concrete progress can actually be made in
understanding primordial star formation. In the case of present-day
star formation, 
we cannot predict  the initial mass function from first
principles despite the wealth of observational data available.
How could we hope to do something like this for unseen primordial stars? 
A scan of the considerable early literature on primordial star
formation  (e.g., Yoneyama 1972; Hutchins 1976; Silk 1977, 1983;
Carlberg 1981; Kashlinsky \& Rees 1983; Palla, Salpeter, \& Stahler 1983;
Carr, Bond, \& Arnett 1984; Couchman \& Rees 1986; Uehara et al. 1996,
Haiman, Thoul, \& Loeb 1996; Omukai \& Nishi 1998)
would tend to support this conclusion. The range of mass estimates
for the first stars spans spans six(!) decades,
from  $1$ to $10^6 M_\odot.$ There are reasons for hope, however.

First, the physics of the first stars is considerably simpler than
that of present-day star formation (Larson 1998; Loeb 1998). The
present-day interstellar medium is an exceedingly
complex environment, but primordial gas initially has no metals, 
no dust grains and no cosmic rays to complicate the gas cooling function. 
Since we think we know the primordial abundances well
and the number of relevant species is small,
the gas chemistry and cooling function are relatively simple and 
have been extensively studied (e.g., see 
Galli \& Palla 1998).
Also, because there are no stars yet, the only relevant external
radiation is the cosmic background whose behavior we also think is
well-understood. Additionally, if one believes that galactic 
strength magnetic fields resulted from dynamo action perhaps enhanced 
by compression (e.g., Kulsrud 1997), then magnetic fields at
early times are likely to be dynamically insignificant.  Finally,
before the first supernovae went off, the early IGM must also have been a 
rather quiescent place, with no sources to sustain turbulent motion.
The remaining consideration in
understanding how primordial gas collapses to form stars
is knowledge of the typical initial conditions: the initial gas density 
and temperature profile, the gas angular momentum distribution,
the density and velocity distribution of the dark matter halos containing
the gas, and the underlying cosmology.  Estimating these
initial conditions for a specific cosmological scenario is no
longer a problem given the current state of 
simulations.  In sum,
at least during the initial phases of primordial star formation, we
have a well-posed problem where the relevant physics is in
hand.

Secondly, computers can now follow the 
inherently three dimensional process of primordial gas fragmentation
and collapse. This is critical because it is not immediately
obvious when gas fragmentation in primordial clouds halts.
To appreciate the difficulties, note that since
gas can cool and increase its density 
arbitrarily (at least until an opacity limit  sets in), 
the Jeans mass for a collapsing gas cloud, i.e., the scale below which 
fragmentation halts, can become extremely small. Such behavior is
indeed seen in one dimensional simulations of isothermal filament collapse. 
One might therefore predict primordial stars to have very low masses.
However, if the initial density perturbations in the cloud are
not very large or the cloud has a very strong central density
concentration, the cloud can collapse into a single object before
the perturbations have time to grow and fragment the cloud
(e.g., Tohline 1980). Thus, depending on exact initial conditions,
one could also predict that the first objects to turn around and
cool will collapse directly into very massive stars or black
holes (the so-called ``VMOs'' or Very Massive Objects).
Furthermore, one cannot straightforwardly apply
the intuition on fragmentation developed in the 
more extensive studies of present-day star formation. 
In the present-day case, gas cooling is very efficient and 
one typically takes the collapsing gas to be isothermal.  
Molecular hydrogen is a very poor coolant, however, and 
the timescale for zero-metal gas to cool can often be 
comparable to or longer than the dynamical timescale for 
the gas to collapse. This has profound consequences, as we show next.

\section{Numerical Simulations of a Collapsing Primordial Gas Cloud}

The results shown are the thesis work of Volker Bromm. (See
Bromm, Coppi, \& Larson 1999 for a more extended discussion
of the calculation.) Our goal here is to 
follow the gravitational fragmentation of a cloud
to see if there is indeed a characteristic mass
scale at which fragmentation stops and gravitational
collapse proceeds unhindered. This ``clump'' mass
scale and the overall spectrum of runaway clump masses that
we find, of course, cannot be directly translated into a 
stellar mass scale or IMF, but it is an important first step.
In the case of present-day star formation, at least,
there is increasing evidence that the two may in
fact be closely related.

Our calculational
approach is intermediate to that of the other 
two primordial gas collapse calculations shown at this meeting. 
The first of these
(see contribution by Nakamura \& Umemura)
uses a high resolution 2-D  mesh code to follow
the evolution of an idealized primordial gas cloud
for many different initial conditions and perturbations.
The second (see contribution by Abel et al.) is a full 3-D
calculation that starts from a large scale cosmological simulation
and uses the adaptive mesh refinement (AMR) technique 
to zoom in on the evolution
of the first object to undergo collapse in their simulation
volume. Only a few realizations of the cosmological 
initial conditions have been explored. In an attempt
to increase the number of realizations we can explore, we
instead follow the cosmological evolution of a tophat density
perturbation with parameters close to those expected
for the first objects that can collapse ($M_{tot}
\sim 10^6 M_\odot,$ e.g., see Tegmark et al. 1997).
We use a 3-D particle code based on the TreeSPH code
of Hernquist \& Katz (1990) that incorporates
the full primordial chemistry of Galli \& Palla (1998),
including the effects of HD cooling. The SPH technique
does not handle shocks as well as mesh techniques,
but strong shocks are not important in the regime considered here
and SPH is simple and flexible. For example, it easy to turn
a group of gas particles into a single ``sink'' particle without
having to worry about mesh artifact/resampling issues. This
is useful for allowing a simulation to continue
beyond  the runaway collapse of the first clump (
which ordinarily would halt the calculation because of
the Courant limit).  It is also easy
to increase our spatial resolution in a desired region 
(i.e., perform a ``poor man's'' version of AMR) by
tagging the particles that enter that region, and then 
restarting the calculation with each of the 
tagged particles replaced by many lower
mass particles, e.g., see Fig. 7.

\begin{figure}[t] 
\plotfiddle{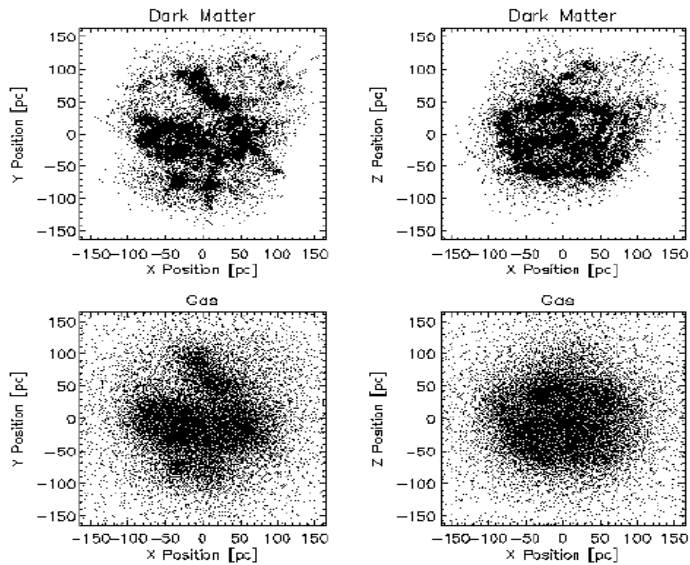}{6.5 cm}{0}{55}{47}{-165}{-90}
\caption{ Projected dark matter and gas distributions at $z=33.5$
The positions of individual SPH particles are plotted.
}\label{z33p5}
\end{figure}

Fig. 1-4 show results one of our typical top-hat collapse/
fragmentation calculations. At $z=100$ we endow
a spherical, uniform density  halo of total mass $2\times 10^{6}M_{\odot}$
(baryonic plus dark matter)  with a Hubble expansion
such that virialization occurs at $z_{vir} \simeq 30.$ 
The dark matter is perturbed with a $P(k) \propto k^{-3}$
power spectrum expected from CDM on small scales. The
baryons are uniformly distributed and have
a mass fraction $\Omega_{B}=0.05.$
Both halo components are initially in solid body rotation about
the $z$ axis, with angular momentum corresponding
to a cosmological spin parameter $\Lambda=0.05.$
These are typical parameters for the first objects
($3 \sigma$ density fluctuation) that turn around 
and are massive enough to cool in a Hubble
time (e.g., see Tegmark et al. 1997). The dark matter initially plays
a key role as the  baryons fall into the potential
wells of the growing small-scale dark matter perturbations (Fig 1).
Eventually, the dark matter undergoes violent relaxation and
starts to lose its substructure. The baryons sink into the center
of the overall dark matter potential well and start to fragment
(Fig. 2). In Fig. 3, we plot the properties of the gas particles
at $z=31.2.$ Note the ``pile up'' of particles at density
$n_H \sim 10^{3-4} {\rm cm}^{-3}$ and temperature $T\sim 300$ K,
corresponding to a Jeans mass $\sim 1000 M_\odot.$
The pile up reflects the fact that gas undergoing collapse
``loiters'' at these values (see the time history in 
Fig. 4, and discussion below).

\begin{figure}[t] 
\plotfiddle{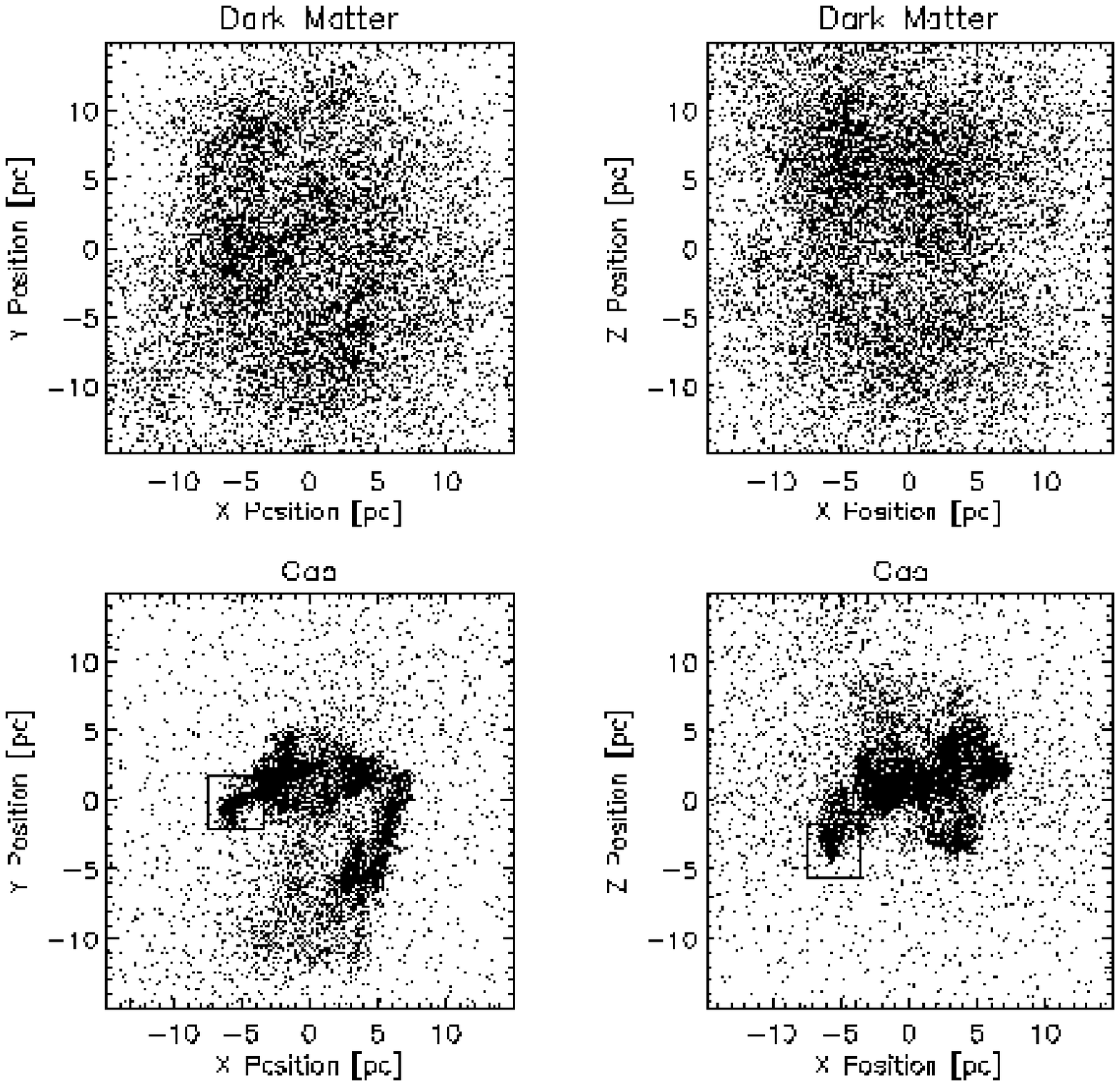}{6.5 cm}{0}{55}{47}{-180}{-87}
\caption{ Dark matter and gas distributions at $z=31.2,$
just before the first gas clump undergoes runaway collapse.
}\label{z31p2}
\end{figure}

\begin{figure}[t] 
\plotfiddle{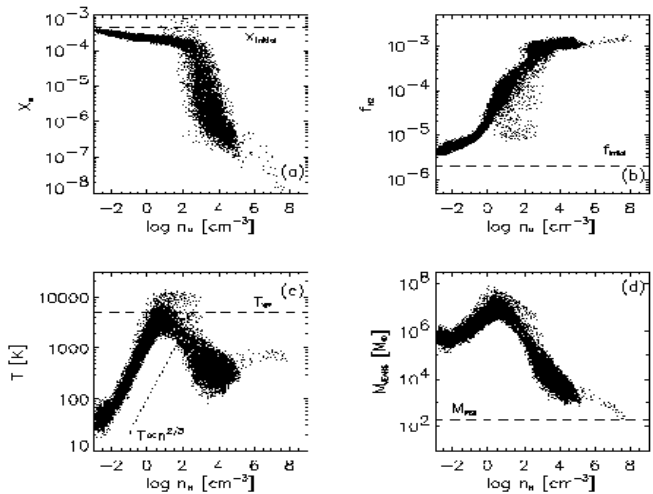}{5.5 cm}{0}{55}{47}{-180}{-105}
\caption{ Gas properties at $z=31.2.$ 
From top left clockwise, we 
plot for each gas particle its (a) free electron abundance, 
(b) fraction of gas in H2 molecules, (c) 
temperature, and (d) Jeans mass corresponding
to the particle's density and temperature, all as functions
of the hydrogen number density in the particle. 
}\label{thermo1}
\end{figure}

\begin{figure}[t] 
\plotfiddle{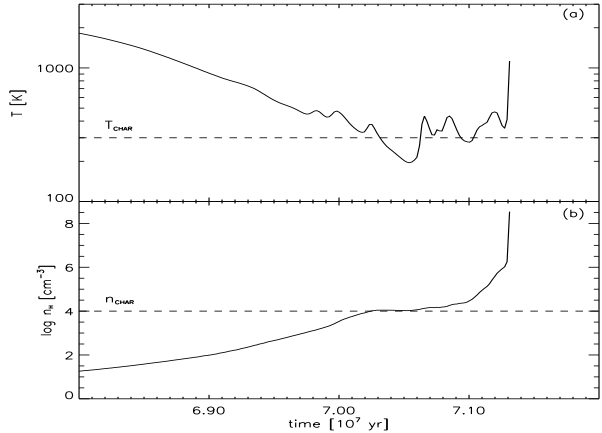}{5.0 cm}{0}{50}{35}{-180}{-100}
\caption{
Temperature and density
time history for the first gas particle in
Fig. 2 to undergo runaway collapse. Note the ``loitering''
phase, $\sim 10^6$ years spent at $n_H\sim 10^4, T=300$ K.
}
\label{tscale}
\end{figure}

\begin{figure}[t] 
\plotfiddle{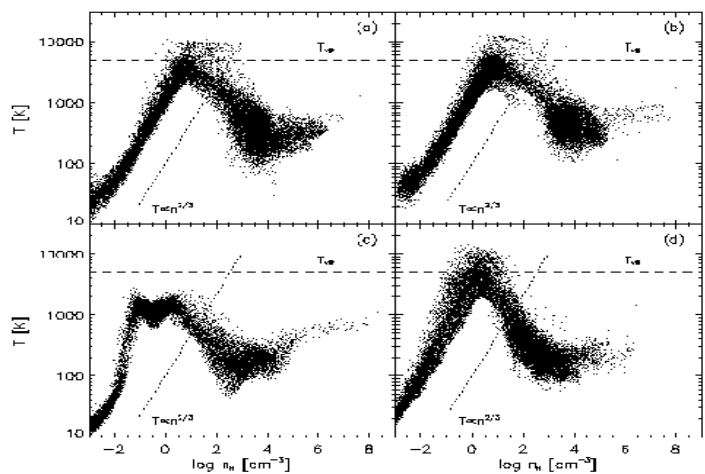}{5.5 cm}{0}{55}{43}{-180}{-90}
\caption{
Thermodynamic behavior (gas particle temperature
vs. density) for  different initial conditions:
(a) the case of Fig.1  but with the number of
gas particles increased by a factor 30, (b) the case of Fig. 1 but with 
a lower angular momentum ($\Lambda=0.02$), (c) a less massive
halo (Fig. 6, $z=27.2$), and (d) a halo
that virializes at $z_{vir}=20.$ Note the clumps at
the same density and temperature as in Fig. 3.
}
\label{manyinit}
\end{figure}

\begin{figure}[t] 
\plotfiddle{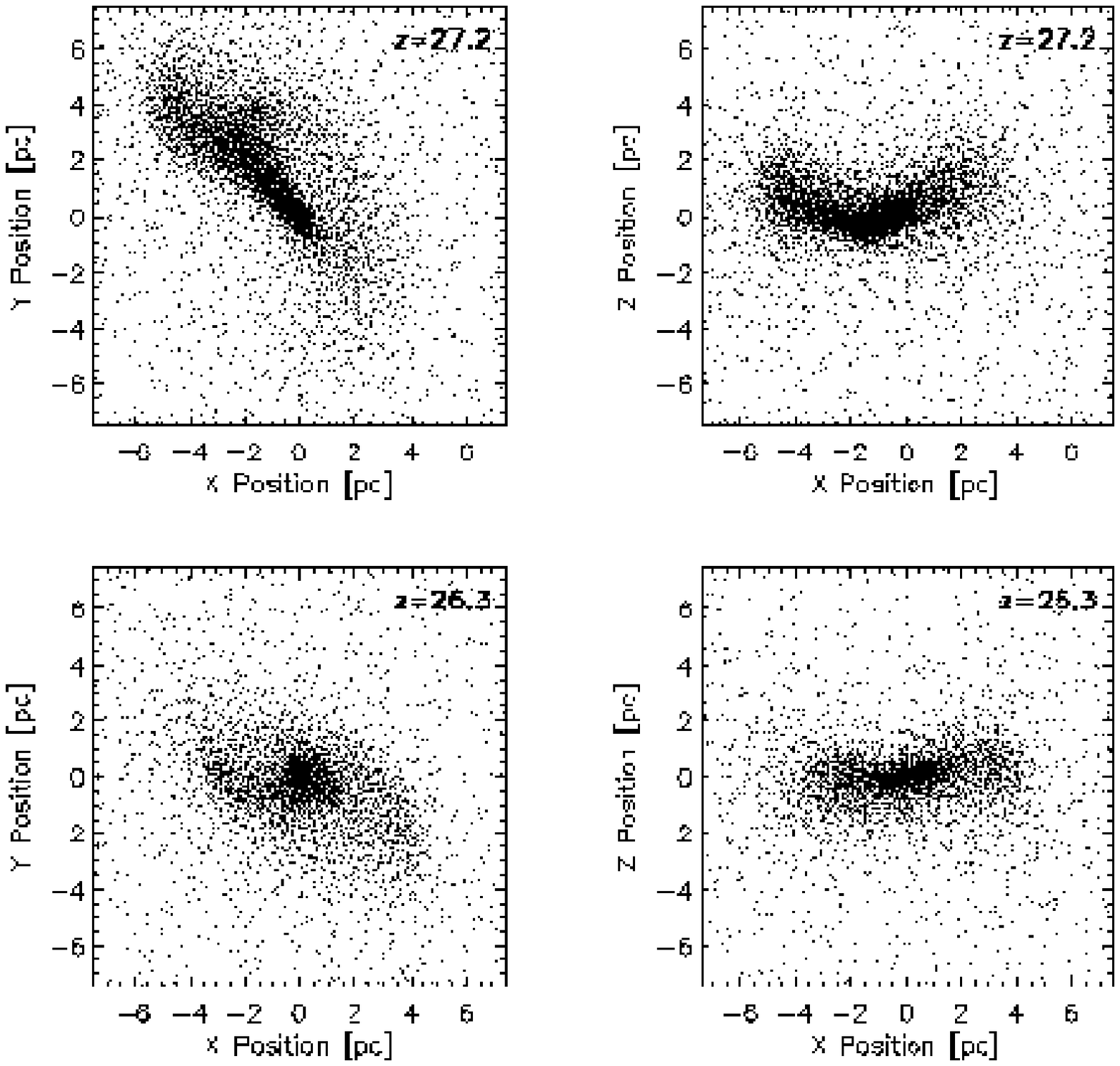}{6 cm}{0}{55}{50}{-180}{-95}
\caption{
Gas morphology at different times
for a low mass halo with $M=2\times10^5 M_\odot$ 
that  marginally satisfies the requirement for efficient
cooling, $t_{cool} < t_{free fall}.$ Only one clump
with final mass $\sim 2000 M_\odot$ forms.
The result is qualitatively
similar to Abel, Bryan, and Norman (1999)
who followed the collapse of such a halo starting from a 
large-scale cosmological simulation.}
\label{manyinit}
\end{figure}

We have carried out many other runs, varying quantities like
the total angular momentum of the cloud, the slope
of the dark matter perturbation spectrum, the degree to which
the mass is centrally concentrated (the standard top-hat assumes
a uniform density distribution, which is optimal in terms
of producing many fragments but may not always be realistic),
the baryon mass fraction ($\Omega_B$), and the total mass and 
turnaround redshift of the cloud.  We find two main results.
First, in terms of the morphology
of the collapsed gas and  the overall ``efficiency'' of fragmentation
(the fraction of gas that ends up in clumps), we find that
varying the initial conditions of the cloud {\it does} make a 
significant difference,
e.g., compare the gas morphology in Fig. 2 with 
that in Fig. 6. Similar dependences, e.g., on the cloud's angular 
momentum and degree of central mass concentration,
are in fact  found in gas simulations
of present-day star formation (e.g., Tsuribe \& Inutsuka
1999).  Note that this dependence on initial conditions
means it is {\it not} possible to make
statements about the overall efficiency of primordial
star formation without first carrying out a comprehensive
survey of the relevant conditions.
Second, despite the differences in gas
morphology,  we always find  
find roughly the {\it same} initial clump masses. Here, 
initial clump mass is defined as the amount of gas that is gravitationally
bound and infalling when the center of a clump starts it runaway collapse,
i.e., it does not include any further gas that may 
eventually accrete onto the clump.

The reason for this perhaps surprising second conclusion
can be found in Fig. 3-5.
If we plot the temperatures and densities of our gas particles when
the first clumps start to collapse, we always find an
excess of particles with temperatures $T \sim$ 200 K and
hydrogen densities $n \sim 10^{3-4}$ cm$^{-3}.$ These two numbers are
not accidental and are set  molecular hydrogen physics -- which
does {\it not} depend on the initial conditions. Specifically,
a temperature  $T\sim 100-200$ K is the minimum one
attainable via H$_{2}$ cooling because of the molecular
energy levels. The corresponding critical density, beyond
which the H$_{2}$ rotational levels are populated according to LTE,
is then $n_{crit}\simeq 10^{3}-10^{4}$ cm$^{-3}$. At the transition
from NLTE to LTE, the cooling rate changes from being proportional
to $n^{2}$ to merely linear in $n,$ i.e., the cooling time required
for the gas to lose a significant fraction of its energy now becomes
independent of density.  Due to this inefficient cooling, the gas `loiters'
and passes through a phase of quasi-hydrostatic, slow contraction
before undergoing runaway collapse (see Fig. 4).
This loitering appears to be crucial as it allows pressure waves to damp
out density anisotropies and inhibits further fragmentation. Although
our results are still somewhat preliminary, we have carried out
higher resolutions runs (e.g., Fig. 7,8) to follow the collapse of 
a clump to much higher densities, and we indeed see no evidence for
sub-fragmentation. Abel et al. have reached the same conclusion
in the even higher resolution runs that they have carried out.
Although we cannot guarantee that some of our clumps will 
not break up into a few objects, e.g., a binary system, it seems
unlikely they will break up into hundreds
or thousands of subclumps. In other words, to astrophysical
accuracy, the Jeans mass $M_J \sim 1000 M_\odot$ that follows from
the typical density and temperature values in Fig. 3 really is
the characteristic clump mass scale for collapsing primordial gas.
The fact that three groups at this conference arrived at the same
conclusion using rather different codes and initial conditions
tells us that a  robust explanation, like the 
physics of molecular hydrogen, must lie behind it.

\section{Implications and Future Directions}
Although we are still far from solving the primordial star formation
problem, the results presented at this meeting indicate we have
made substantial progress. Unless our understanding of primordial
gas cooling is very wrong (note that our simulations included HD 
cooling which some have speculated to be important)  or the typical physical 
conditions during the early dark ages are very different from 
current expectations, it appears inescapable
that the first typical objects to collapse will fragment into clumps
of initial mass $\sim 1000 M_\odot.$
It also appears likely that these clumps will not
fragment much further as their interiors collapse to form a 
star or a few stars. 
Therefore, typical primordial protostars
are likely to look quite different from those around us today,
and in particular, will be 
surrounded by much more massive $\sim 100-1000 M_\odot$ envelopes.
As noted at this meeting, it is not at all
obvious how much of this mass actually makes it onto the final 
star. However, given that all the scales are so much larger
and resemble those we find around present-day massive
stars in the process of being born, it is difficult
to see how one can make ordinary solar mass stars from
such a gas configuration, i.e., primordial star formation
is probably strongly biased towards massive ($\sim 10-100 M_\odot$) 
and possibly very massive ($\sim 1000 M_\odot$) stars.
This would explain why we see no zero-metal stars today
and has important consequences that have not been fully
explored yet, e.g., 
massive primordial stars produce many more ionizing UV photons per unit
mass than low mass ones 
(see Bromm, Kudritski, \& Loeb 2000 for a detailed calculation
of the spectrum from a massive zero metal star).Also, such 
massive stars could be good progenitors for hypernovae and gamma-ray bursts,
or the seeds for massive black hole formation.

There remain important questions that do not require
a major leap in computing power to answer. First, at this meeting
it became clear we need to deide what are typical, realistic initial 
conditions, e.g, Nakamura \& Umemura pointed out that it is 
possible to fragment down to $\sim 1 M_\odot$ if one can start out with
dense enough filaments ($n_H \ga 10^6 {\rm cm}^{-3}).$
We agree since this would skip the ``loitering'' phase of the 
collapse, but we do not see how such filaments arise in a realistic
scenario. Quantifying the relevant initial conditions for primordial
gas collapse will let us determine the efficiency of clump formation,
which in turn gives us an upper
limit on the primordial star formation efficiency and
provides a first indication as to the importance of the first
stars. Secondly, we can consider what happens to  gas
collapse and fragmentation in the presence of trace metals and
UV background radiation from a previous generation of stars.
Population III
star formation is often considered to be a short-lived
event because the pristine conditions required are wiped out
once the first stars produce UV light and the first supernovae
produce metals. However,
as metallicity does not build
up instantly, there may well be an extended window
of time when star formation either proceeds in the massive
clump/star mode described here (if H2 is present) or 
not at all (if H2 is destroyed by UV radation).  
Our preliminary calculations indicate that metal cooling does
not become important until the gas metallicity reaches
$Z\ga 10^{-4}-10^{-3} Z_\odot$ -- which coincidentally is the
range of the lowest observed metallicities  and also the 
range where abundance anomalies begin to appear in metal poor
stars. (These anomalies are often interpreted as increased scatter
due to enrichment by individual supernova events, but they could
also reflect atypical progenitor stars, e.g., that had much hotter
interiors than stars  today.) 
Finally, it should be
possible to push the spherically symmetric protostar 
calculation of Omukai and Nishi (1998) through to the accretion
phase (along the lines of Masunaga, Miyama, and Inutskuka 1999
in the present-day star formation case). This will enable a first cut at 
understanding the feedback of the primordial protostar's radiation 
on its envelope. The feedback is likely to be strong, and Abel
conjectured at this meeting, for example, that all accretion may 
stop once the protostar 
reaches $\sim 10 M_\odot$ and produces enough ionizing
radiation to destroy the envelope's molecular hydrogen,
thereby removing its primary means of cooling.
Without a real calculation, however, it is not clear what the 
outcome will be. If the envelope gas  simply becomes adiabatic,
accretion can still occur if a sufficient central mass
concentration has already been established (Bondi 1952).

\begin{figure}[t] 
\plotfiddle{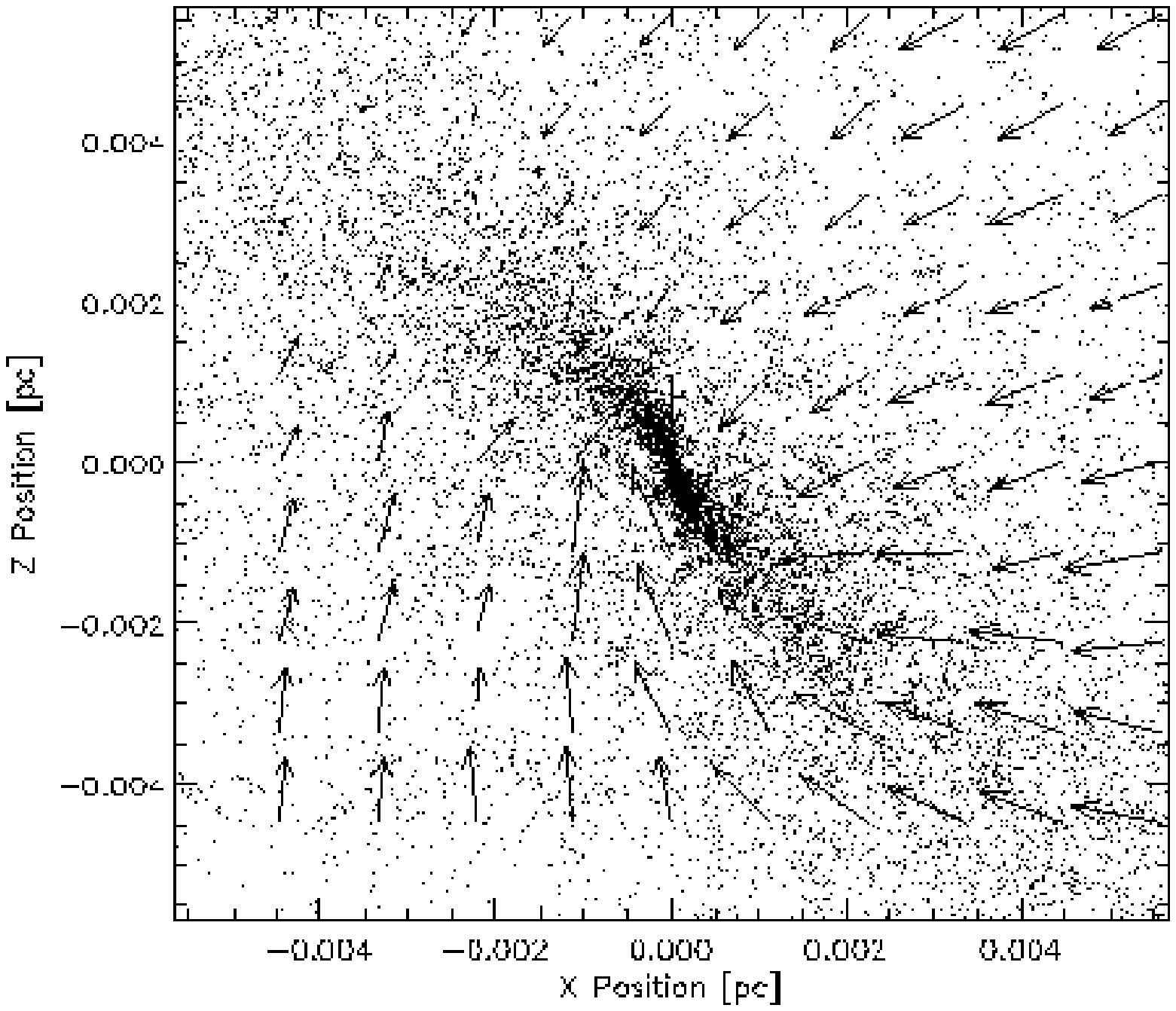}{5.5 cm}{0}{55}{48}{-170}{-110}
\caption{
Gas morphology and kinematics near 
density maximum of first clump to form in 
Fig. 1. The number of particles
has been increased by a factor 100, and the linear size of 
the plot is now $\sim 2500$ AU. The longest
arrow represents a velocity of $14.5$ km/s. The spindle structure
at the center has {\it not} fragmented and contains a few tens of 
solar masses. The surrounding
flow is supersonic with Mach numbers $M \sim 3-5.$
}\label{resample}
\end{figure}

\begin{figure}[t] 
\plotfiddle{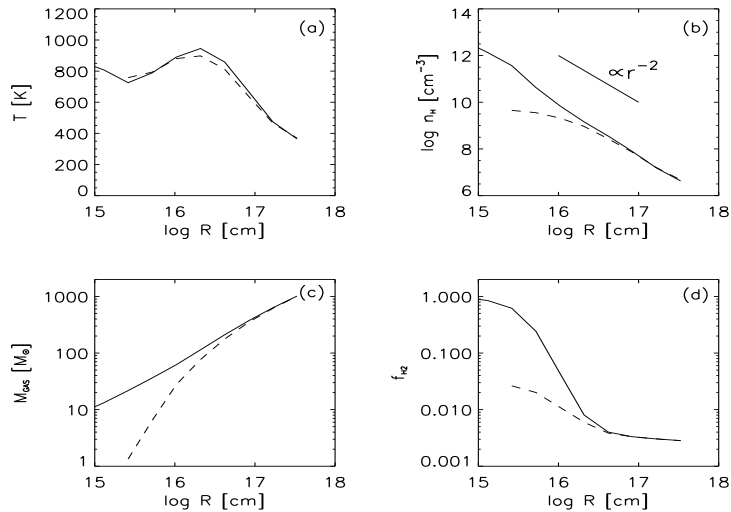}{5.5 cm}{0}{60}{48}{-180}{-155}
\caption{
{\it Solid lines: } 
mass averaged structure of the clump in Fig. 7
vs. radial distance from the clump density maximum.
{\it Dashed lines:} 
same, but $\sim 1000$ years earlier.
{\bf (a)} Gas temperature.
{\bf (b)} Hydrogen number density.
An extended envelope forms with an approximately isothermal
density profile, $\rho \propto r^{-2}.$
{\bf (c)} Enclosed gas mass.
A $\sim 100M_{\odot}$ core begins to freely fall,
while the rest of the heavy envelope 
hardly moves on the evolutionary timescale $\sim 10^{3}$ yr.
{\bf (d)} H$_{2}$ fraction. At 
$r < 10^{16}$ cm, three-body reactions convert
the hydrogen into almost fully molecular form.   
}\label{resample}
\end{figure}


\bigskip
\noindent $^*{\rm Our}$ work was supported by NASA grant NAG5-7074. 
We thank the organizers
for a well-run and stimulating conference.

\end{document}